# Encoding many channels in the same frequency through radio vorticity: first experimental test.

Fabrizio Tamburini[1], Elettra Mari[2], Anna Sponselli[1], Filippo Romanato[3,4], Bo Thidé[5], Antonio Bianchini[1], Luca Palmieri[6], Carlo G. Someda[6].

[1]Department of Astronomy, University of Padova, vicolo dell'Osservatorio 3, I-35122 Padova, Italy, EU.

[2] CISAS, University of Padova, via Venezia 15, I-35131 Padova, Italy, EU.

[3] Department of Physics, University of Padova, via Marzolo 8 I-35100 Padova, Italy, EU.

[4] LaNN, Laboratory for Nanofabrication of Nanodevices, Venetonanotech, via Stati Uniti 4, IT-35100 Padova, Italy, EU.

[5]Swedish Institute of Space Physics, Box 537, Ångström Laboratory, SE-75121 Uppsala, Sweden, EU.

[6]Department of Information Engineering, University of Padova, via Gradenigo 6/B I-35131 Padova, Italy, EU.

E-mail: fabrizio.tamburini@unipd.it

**Abstract.** We have shown experimentally that it is possible to propagate and use the properties of twisted non-monochromatic incoherent radio waves to simultaneously transmit to infinity more radio channels on the same frequency band by encoding them in different orbital angular momentum states. This novel radio technique allows the implementation of, at least in principle, an infinite number of channels on one and the same frequency, even without using polarization or dense coding techniques. An optimal combination of all these physical properties and techniques represents a solution for the problem of radio band congestion. Our experimental findings show that the vorticity of each twisted electromagnetic wave is preserved after the propagation, paving the way for entirely new paradigms in radio communication protocols. **Pacs (41.20.-q, 07.57.-c, 41.20.Jb, 42.60.Jf)**

## Contents:

# Introduction

The first radio signal transmitted and received by Guglielmo Marconi on 8 December 1895 started the revolution of wireless communication in our modern world[1]. Nowadays, information is mostly exchanged through wireless channels and the rapid increase of the use of mobile devices has led to a congestion of the available radio bands even after the application of dense coding and channel sharing techniques[2].

We report experimental results that demonstrate a new revolutionary method of wireless communication that characterizes and exploits the propagation in the far field zone of orbital angular momentum (OAM) states[3] of general electromagnetic waves, away from the paraxial approximation. Our findings extend the previous indoor laboratory test where the transmission of optical twisted OAM states of coherent paraxial laser beams is discussed[4] and show that vorticity is preserved after the wave propagation.

OAM states are connected with the momentum carried, together with energy, by the EM field. The linear momentum is connected with force action and the total angular momentum, $J = S + L$, with torque action. The quantity $S$ represents the spin angular momentum, related to photon helicity and thus with polarization. The term $L$ is associated with the orbital helicoidal phase profile of the beam in the direction orthogonal to the propagation axis, obtained through the discrete superposition of photon eigenstates, each with a well-defined value $\ell\hbar$ of OAM[5,6,7,8]. When a paraxial beam of light propagates in vacuum along the $z$ axis, one can project $S$ and $L$ onto $z$ and obtain two distinct and commuting operators, $S_z$ and $L_z$[9]. The amplitude of the EM field measured in the plane orthogonal to $z$, $U_{\ell, p}^{LG}$, is then described in terms of Laguerre-Gaussian modes,

$$U_{\ell,p}^{\angle-G}(r,\vartheta) \propto \left(\frac{r\sqrt{2}}{w}\right)^{|\ell|} L_p^\ell\left(-\frac{r^2}{w^2}\right)\exp\left(-\frac{r^2}{w^2}\right)\exp(-i\ell\vartheta) \qquad \text{(eq. 2)}$$

where $\ell$ describes the number of twists of the helical wavefront, $p$ the number of radial nodes of the mode, $w$ the beam waist, $L_p^\ell(x)$ is the associated Laguerre polynomial, and $\vartheta$ is the phase. OAM finds practical applications in many fields: radar[25], nanotechnology[10], quantum experiments[11]; in astronomy[12,13,14,15,16] to improve the resolving power of diffraction-limited optical instruments[17], facilitate the detection of extrasolar planets[18] and of Kerr black holes[24].

## Transmitting with radio vortices

Up to now, no one had managed to do a real radio transmission of twisted beams outdoor, in the real world, by using an incoherent achromatic wide band of twisted radio beams trapped inside a generic antenna lobe. We prove experimentally that on one and the same frequency, the natural orthogonality of these states provide ideally an infinite set of independent transmission channels, each characterized by their unique spatial phase front topological characteristic. To this aim we set up an experiment, where we generated and detected two different channels on the same frequency band: the first untwisted, with $\ell = 0$ orbital angular momentum, and the other with an $\ell = 1$ OAM twist. Two equal transmitters, each with the power of 2 Watts, tuned to the same frequency of $\lambda = 2.414$ GHz were feeding the antennas. The bandwidth used in the transmission was 15 MHz, like that used to transmit video signals. The $\ell = 0$ source was generated by a commercial 16.5 dBi Yagi-Uda antenna[19]. To generate the $\ell = 1$ vortex beam, we modified mechanically a 26 dBi commercial off-axis parabolic antenna to attain an off-axis spiral parabolic shape.

## Mapping of the field

As first step, we experimentally characterized the physical properties of the twisted uncorrelated achromatic EM wave train, proving that vorticity can be actually radiated all the way to infinity[20] and that the topological properties of twisted waves, namely, the presence of the singularity and the spatial phase signature are preserved in the far-field zone, even far away from the usual paraxial approximation used for laser beams. The intensity distribution of the radio vortex was mapped 40 metres distant from the transmitting antenna ( 320 $\lambda$ ). The average signal background measured at $\lambda = 2.414$ GHz inside the 15 MHz bandwidth was -90 dBm. The polarization was kept horizontal. The expected beam waist was given by the diffraction limit of the antenna, $\delta\varphi = 1.22\dfrac{\lambda}{D} \approx 10.9^0$ and the half-power beam width (HPBW), the angular separation between the points on the antenna radiation pattern at which the power drops to one-half (-3 dB) its maximum value, was found to be $\theta = k\lambda / D = 8.75^0$. The HPBW diameter at 40 metres distance is on the order of 6 meters. We found that the region harbouring the 3 cm wide vortex singularity appears to be a widely structured dip with a size of 21cm, as shown in figure 1. The intensity pattern of the main lobe and the phase fingerprints of the beam without OAM, instead, were almost constant in the plane orthogonal to the emission direction of the antenna.

## Radio transmission with OAM

Differently from the already existing artificial protocols that use the spatial phase distribution generated by a set of antennas, the advantage provided by OAM is to have a precise, natural and stable phase fingerprint that can be used as reference pattern to tune each of the vortex channels in phase. These states of the EM field have been proved to be independent orthogonal physical states and can be generated, modulated and propagated independently without causing mutual interference between each other. EM-OAM is a fundamental physical property of the EM field that, applied to wireless communication, can offer additional degrees of freedom to use as a set of new orthogonal and independent communication channels at any given frequency become available. Our findings prove that this property can be used for long range communications.

After having verified that the properties of the twisted beam were preserved, we transmitted and received the two channels in the directory from the lighthouse of San Giorgio island to the balcony of Palazzo Ducale in Venice, separated by 442 metres ($3536\,\lambda$). To better align the receiver with the transmitting stations we adopted the find-and-track direction method[21]. The HPBW diameter was 66 meters and the darker zone where the two-wavelengths wide central singularity is harboured was ~190 cm. We identified and measured the two radio signals, twisted and untwisted, by using an interferometric phase discrimination method that puts in evidence the phase fingerprints due to the twisting of electromagnetic waves[22,23,24,25]. The receiving station consisted of a receiving frequency-modulation radio module fed by two identical 16.5 dBi Yagi-Uda antennas (hereafter antenna **A** and **B**) connected together with 180° dephased cables through a beam adder module in order to obtain a phase-difference interferometer. The interferometer **A-B** detects the spatial phase properties of the beams, fingerprints of their different vorticity states. Antenna **A** was mounted on a mechanical translator oriented towards the direction of the transmitting station to select one of the two channels by exploiting the spatial phase front properties of different OAM states present in the two beams, whereas antenna **B** could be adjusted in the orthogonal direction. The phase difference $\phi$, acquired by EM waves propagating along the two paths from the source to the two receiving antennas **A** and **B**, is

$$\phi = 2\pi \frac{d \sin\theta}{\lambda},$$

where $d$ is the separation of the phase centers of the two antennas, $\lambda$ is the radio wavelength and $\theta$ the polar angle. The signal is collected equally by antenna **A** and **B** in phase and the signal of antenna **A** arrives at the signal adder 180° out of phase with respect that of antenna **B** because of the electric $\lambda/2$ cable delay, resulting in a difference signal configuration, $|\mathbf{A} - \mathbf{B}|$ such that

$$\left| A - B \right| \approx \left| V_0 - V_0 e^{i\phi} \right| = 2V_0 \sin\frac{\phi}{2}$$

where $V_0$ is the voltage measured by the receiver at the antenna cable end. The direction of the transmitter, in the ideal case, is identified by a minimum or absence of signal. A maximum is obtained when $\phi = (k+1)\pi$ and $k$ is an integer.

By adding a phase term to **A**, one can change the pointing direction of the antenna system or moves the position of the null interference fringes. Alternatively, by moving the antenna **A** along the direction of the source a quantity $\Delta x = 2\pi n/\lambda$, the phase difference between the two paths becomes

$$\phi = 2\pi \frac{d\sin\theta - n}{\lambda} \; .$$

The parameter $n$ can be adjusted to improve the tuning of the receiving system and read a signal minimum in the exact direction of the transmitting antenna. Here, $n$ is negative when the antenna is moved towards the source.

If the beam carries OAM $\ell \neq 0$, the phase distribution of the wavefront arriving to antenna **A** and **B** will exhibit a characteristic topological signature. In the simplest case, when the centre of the vortex coincides with that of the interferometer, the two antennas will experience a phase gap due to the OAM of the EM wave $\phi_{OAM} = \ell\pi$ and a maximum of the signal is obtained when the phase factor is

$$\phi = 2\pi \frac{d\sin\theta - n}{\lambda} + \ell\pi = (k+1)\pi; \quad k \in \mathbb{Z} .$$

A maximum for the vortex is achieved when $n = 0$ and $k = 0$. Because of destructive interference, at the same time, the $\ell = 0$ signal intensity will experience a minimum. On the other hand, a maximum, for the $\ell = 0$ mode will be found when $n = -\lambda/2$, corresponding to a minimum for the vortex.

Following these considerations, we aligned the interferometer on the center of the singularity so to obtain the phase gap of $\pi$ between the two antennas of the $\ell = 1$ vortex and the baseline was oriented by an inclination of ~10° with respect to the balcony to be orthogonal to the beam. We transmitted simultaneously two constant audio signals at different tones for each OAM channel on the same frequency carrier (440 Hz and 1000 Hz for the planar and twisted wave, respectively). On a single-antenna receiver the two radio signals were audible superposed together. By using the interferometer and mechanically delaying the antenna **A** with respect to **B** to select one of the two orthogonal OAM beams, one signal is alternately shut off with respect to the other because they carried different OAM states and hence a different spatial phase signature. As shown in figure 2, we demonstrated an excellent discrimination between the two different OAM channels: the $\ell = 0$ beam shows a wide minimum of the carrier in the 12 cm translation of the antenna, as expected, while the twisted beam shows a rich forest of minima due to its spatial topological properties sensed by the Yagi lobes. In the superposition of the two signals the transition between the $\ell = 0$ and $\ell = 1$ OAM modes is characterized by a complicated pattern of the two alternating signals because of the secondary Yagi lobes, with regions in which were audible either the twister channel or the untwisted one, as shown in the latest panel of figure 2. An audio recording, divided in three mp3 files of the tuning between the two OAM channels is provided as additional material. The first audio file, l0.mp3, is the recording of the spatial tuning of the channel without OAM. One can hear the main tone at 440 Hz and then a strong white noise in the position where antenna **A**, moving in the direction of the source with respect to antenna **B**, reaches the point where the signal is cancelled by the interferometer. Two minor bumps are seen in correspondence to the secondary Yagi lobes. The second file l1.mp3, instead, shows that the twisted beam has a much richer spatial structure than that of the twisted beam. Finally, the third file is the recording of the vortex tuning between the two different OAM states sent simultaneously on the same frequency.

Already with this setup one can obtain four physically distinct channels on the same frequency by additionally introducing the use of polarization, in this case independent from OAM. A further multiplication of a factor five after the implementation of multiplexing, yields a total of 20 channels in the same frequency.

## Conclusions

Our results open new perspectives in wireless communications and demonstrate the possibility of tuning along different orbital angular momentum channels without increasing the frequency bandwidth from a physical principle. Looking back to the past, history books report that Marconi invented wireless telegraph and from that the communication world spread its branches in all directions[1]. All the wireless devices are based on various forms of phase, frequency and/or amplitude modulation of the electromagnetic (EM) radiation. In order that many different broadcasting stations could transmit simultaneously without overlapping their radio signals, Marconi suggested a division of the available spectrum of radio frequencies in different bands[22]. Nowadays, the wide use of wireless has unavoidably led to the saturation of all available frequency bands, even with the adoption of artificial techniques to increase the band capacity. With this experiment we have shown that the use of orbital angular momentum states dramatically increases the capacity of any frequency band, allowing the use of dense coding techniques in each of these new vortex radio channels. This represents a concrete proposal for the solution of the band saturation problem.

Moreover, with this experiment we verified the propagation of the physical properties of twisted radio beams. We found that the spatial phase signature was preserved even in the far-field region and for incoherent non-monochromatic wave beams. These results open new concrete perspectives for science, including astronomy, radio-astronomy, with striking results for the detection of Kerr black holes in the test general relativity[24].

## Acknowledgments

The authors acknowledge the logistic support of the Department of Engineering Information, University of Padova and CNR in the building and testing of the setup, the "Vortici & Frequenze Group", Tullio Cardona, and Venezia Marketing Eventi for the logistic support during the experiments in Venice "Onde sulle Onde". We also acknowledge the support of ARI - Venice (Italian Radio Amateur association), Fondazione Marconi, Princess Elettra Marconi and Dr. Ing. Giuliano Berretta, Eutelsat. BT acknowledges the University of Padova for support and hospitality. The experiment has been partially funded by University of Padova project "Study of angular orbital momentum".

**Figure 1**

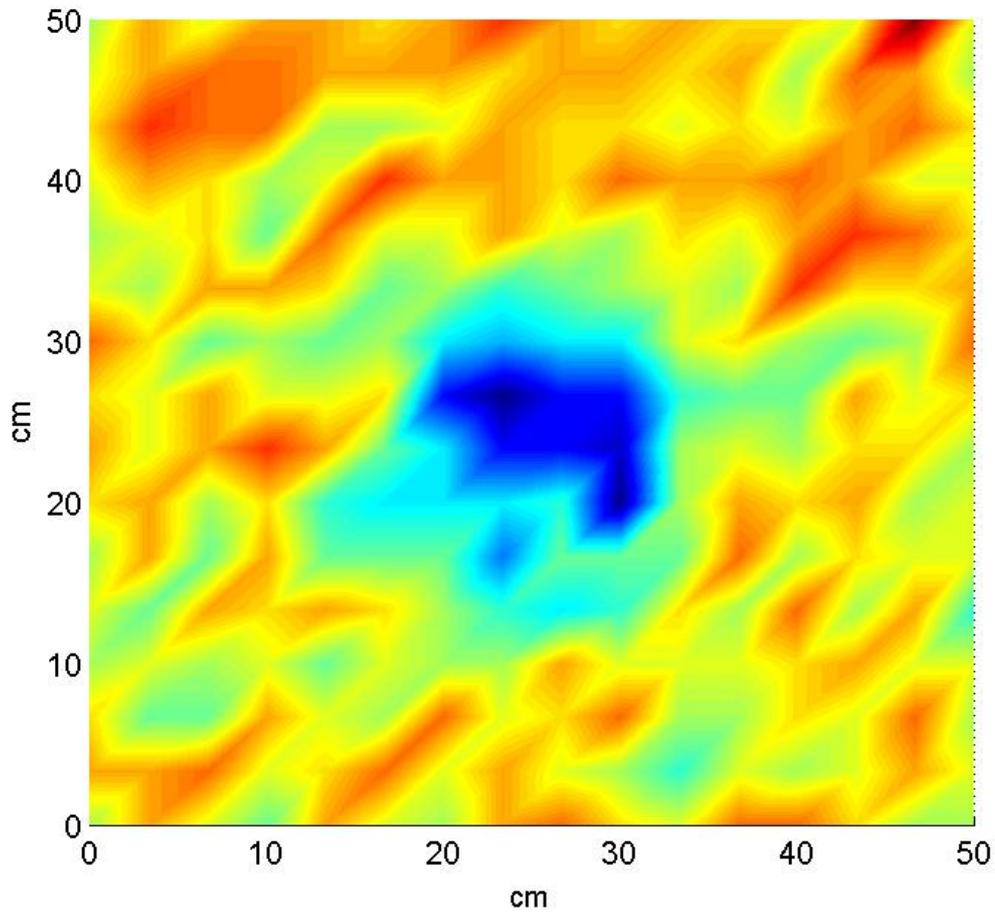

Figure 1: Intensity map of the radio beam vortex at 40 metres in open space in the region harbouring the singularity. The intensity distribution in this region presents some fluctuations caused by environmental interference effects. The central dip indicates the region where is located the field singularity, less than 3 cm wide, with measured intensity -82 dBm. The actual position of the singularity was then confirmed by the phase change measured by the two-antenna interferometer A-B (see text). The scale is in centimetres.

**Figure 2**

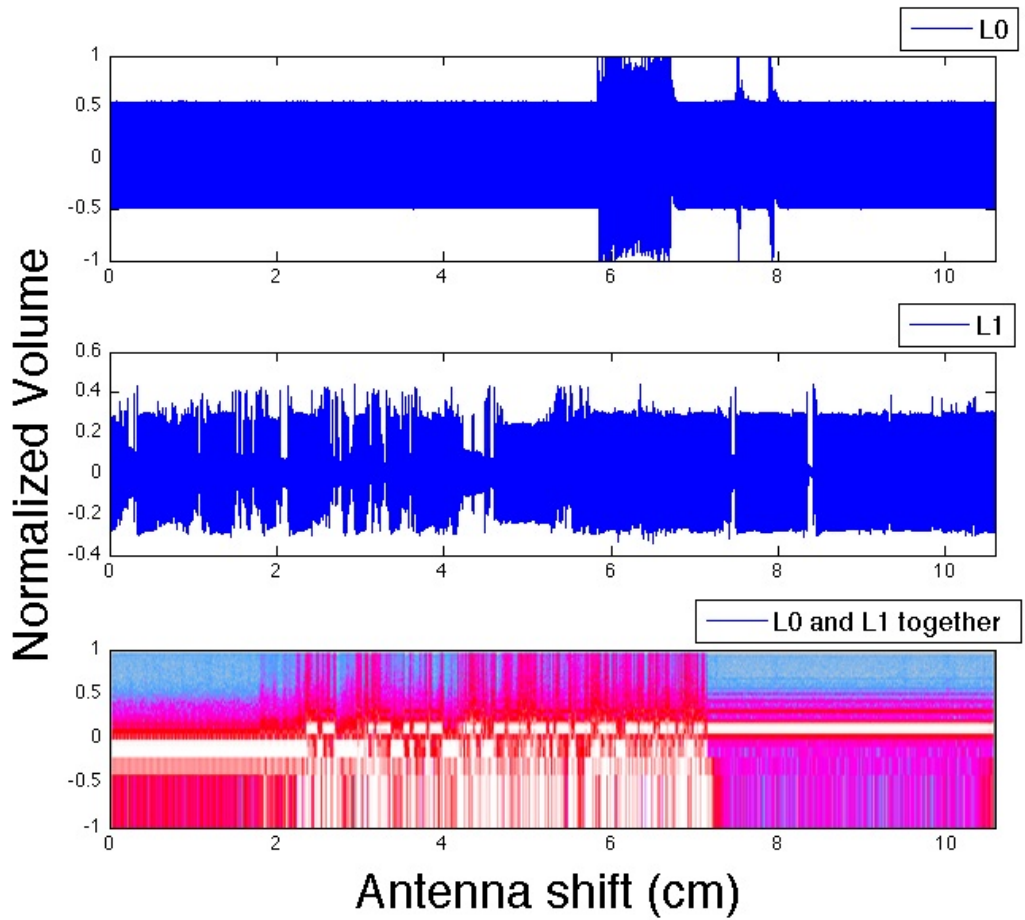

Figure 2: diagram of the monophonic audio recordings with 32 bit precision. Top: the untwisted wave shows a region of null signal, in which the loud white noise (5.8 - 6.7 cm antenna shift) replaces the constant modulation in the carrier, with two minor regions due to the secondary Yagi lobes at the positions 7.5 and 7.95 cm. The 440 Hz sound is replaced by white noise at 6 cm. Middle panel: the $\ell = 1$ twisted beam shows a wide forest of maxima and minima because of the spatial pattern of the vortex read by the two Yagi lobe structure. Bottom panel: Audio tracking of the superposition of the two signals. The transition from the 440 Hz carried by the $\ell = 0$ mode to the 1 KHz sound of the $\ell = 1$ mode is clearly visible. In the middle a transition between the two modes caused by the secondary lobes of the Yagi antenna leaving the vortex region is seen. The bandwidth was 15MHz wide. Each dataset is 22870008 bytes long. The corresponding audio files, in mp3 format, are included as additional material.

## Appendix:

### *Waves on Waves: "Onde sulle onde", a public experiment*

The final results of this experiment were publicly shown in Piazza San Marco, Venice, on June 24 2011 to the international press and citizenship. In the style of Guglielmo Marconi, we realized the first public demonstration of the radio vortex performances to involve also common people in this experiment. A different way of communicating science. A light and sound show, projected on Palazzo Ducale explained people what the experimenters were doing. More than two thousand people attended the "live" experiment at 21.30 local time and when the signal was tuned from zero to vorticity one in the same frequency and transmitting simultaneously, a rifle shot was heard. This to remember the first radio transmission made by Guglielmo Marconi in 1895. And in the facade of Palazzo Ducale was projected "Segnale Ricevuto", signal received. Many videos can be found in youtube.

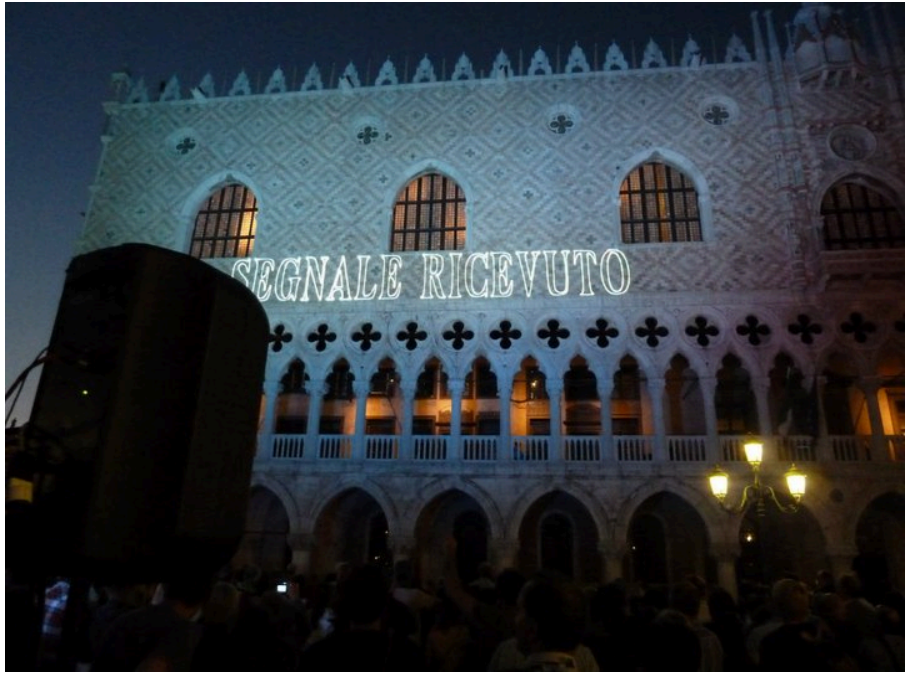

**Figure 1A: Twisted signal characterized and received (segnale ricevuto).**

## *Hardware*

To demonstrate the feasibility of implementing multiple radio communication channels in the same frequency, discriminated only by OAM states, we decided to adopt a very basic hardware configuration, made by a couple of commercial audio/video tunable 2 Watt transmitter and receiver modules feeding the twisted parabolic antenna for the twisted beam with OAM, and a Yagi antenna for beams without OAM.

The receiver is made from two identical Yagi antennas tuned at 2.4 GHz, mounted on the top of two identical plastic columns ($\varepsilon = \varepsilon_0$). The baseline was of 4.50 meters, with laser controlled levering and calibrated mutual distance. Each of the receiving antennas were mounted on a translator that provides fine tuning in an interval of 10 cm ( $\sim 0.8\lambda$). The connection cables were made with Belden H155, 50 Ohm impedance, diameter 5.4 mm. The velocity factor is 79% with a dB loss/100m @ 100MHz 9.3dB. The half-wavelength junction for the phase difference interferometer is 4.94 cm.

### Transmitter/receiver modules:

The audio and radio signals for the calibration and radio transmission were generated with commercial high-quality super-heterodyne frequency modulating (FM) transmitting modules. The changes in frequency due to FM modulation of the carrier can be considered negligible and do not cause a significant distortion of the vortex or a consequent change of the topological charge obtained with the twisted antenna, since the modulation frequency is up to two orders of magnitude smaller than the carrier frequency. In the $\ell = 0$ beam a sinusoidal audio modulation with the frequency of 440 Hz was used, while in the twisted channel was modulated at 1 KHz, as shown in the power spectrum of the audio recordings in the figure below.

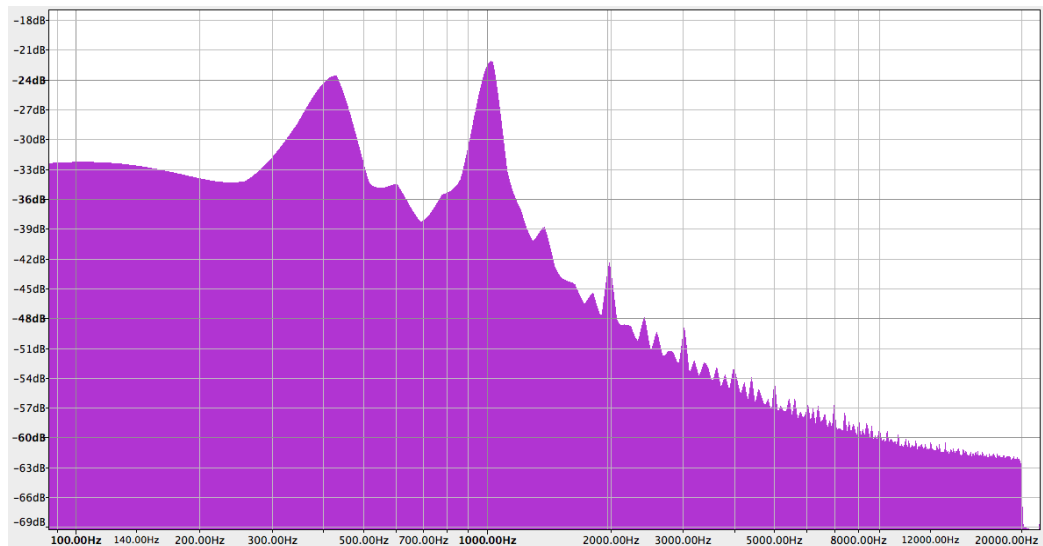

**Figure 2A: Power spectrum of the audio signals during the superposition of the two OAM channels. The two first peaks correspond to 440 Hz of the $\ell = 0$ mode and 1 KHz of the $\ell = 1$ mode. The frequency scale is logarithmic.**

## Offset vortex parabolic antenna:

From a pair of equal 80 cm, 15° offset, steel parabolic antennas, the dish was transformed into a vortex reflector by elevating, from the original shape, the surface of the quantities reported in the table for given values of the azimuthal angle

| Azimuthal angle | Elevation (in units of λ) | Elevation (cm) |
|---|---|---|
| 0=2π | 1/2 | 6.25 |
| π/2 | 3/4 | 4.69 |
| π | 1/4 | 3.12 |
| 3/2 π | 1/8 | 1.56 |

The expected lobe of the parabolic antenna is

| dB contour down | 0 | 0.25 | 0.5 | 1 | 24 | |
|---|---|---|---|---|---|---|
| Full beamwidth degrees | 0 | 3.13 | 4.42 | 6.25 | 8.84 | 10.83 |
| Gain dBi | 24.2 | 23.95 | 23.7 | 23.2 | 22.2 | 21.2 |

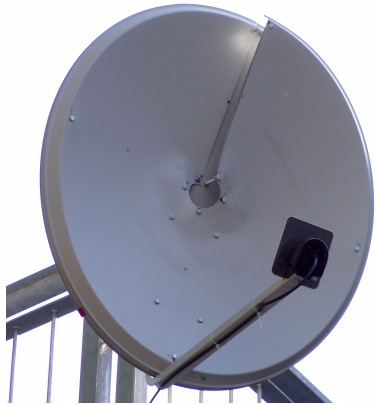

**Figure 3A: the twisted parabolic antenna**

## Site of San Marco experiment

**Transmitter        T:**                          **Receiving station      R:**
**Torretta della Compagnia della**              **Loggetta del Palazzo Ducale**
**Vela in San Giorgio (lighthouse)**

Altitude 10m msl                                 Altitude 16m msl

lat      45°25' 48" N                            lat      45°26'00" N

long    12°20' 35" E                             long    12°20'25" E

Geometrical distance 442 meters c.a. measured with GPS.

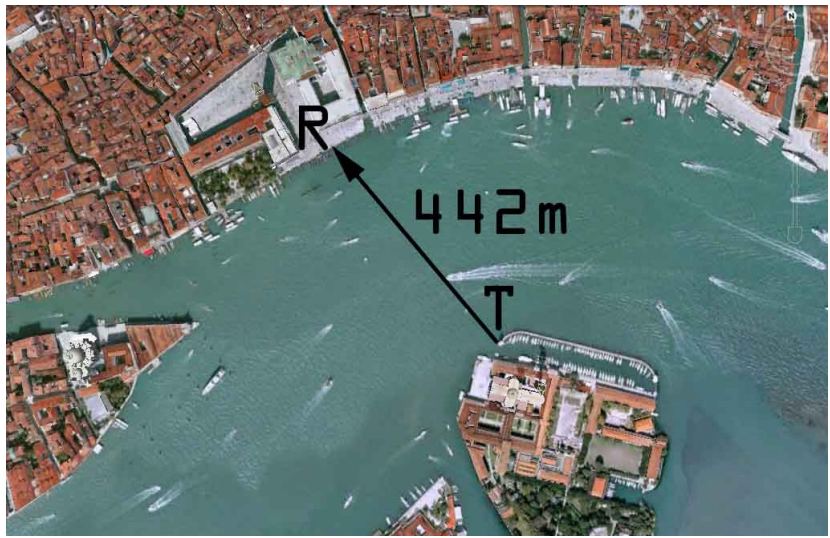

**Figure 4A: Map of the experiment site in San Marco, (Italy)**